\documentclass[prl,twocolumn]{revtex4-1}
\usepackage{amsmath,amsfonts}    
\usepackage{graphicx}   
\usepackage{color}
\usepackage{hyperref}  
\usepackage{epstopdf}
\usepackage{subfig}
\usepackage{rotating}
\DeclareGraphicsExtensions{.pdf,.jpeg,.png}

\providecommand{\abs}[1]{\ensuremath{\left| #1 \right|}}

\providecommand{\lap}{\ensuremath{\nabla^2}}

\begin{document}
\title{Attraction, Merger, Reflection, and Annihilation in Magnetic
  Droplet Soliton Scattering}
\author{M. D. Maiden}
\affiliation{Deparment of Mathematics, Meredith College, Raleigh, North Carolina  27607, USA}
\author{L. D. Bookman}
\author{M. A. Hoefer}
\email{mahoefer@ncsu.edu}
\affiliation{Department of Mathematics, North Carolina State University, Raleigh, North Carolina 27695, USA}
\begin{abstract}
  The interaction behavior of solitons are defining characteristics of
  these nonlinear, coherent structures.  Due to recent experimental
  observations, thin ferromagnetic films offer a promising medium in
  which to study the scattering properties of two-dimensional magnetic
  droplet solitons, particle-like, precessing dipoles.  Here, a rich
  set of two-droplet interaction behaviors are classified through
  micromagnetic simulations.  Repulsive and attractive interaction
  dynamics are generically determined by the relative phase and
  speeds of the two droplets and can be classified into four types:
  (1) merger into a breather bound state, (2) counterpropagation
  trapped along the axis of symmetry, (3) reflection, and (4) violent
  droplet annihilation into spin wave radiation and a breather.
  Utilizing a nonlinear method of images, it is demonstrated that
  these dynamics describe repulsive/attractive scattering of a single
  droplet off of a magnetic boundary with pinned/free spin boundary
  conditions, respectively.  These results explain the mechanism by
  which propagating and stationary droplets can be stabilized in a
  confined ferromagnet.
\end{abstract}

\maketitle

Solitary waves or solitons are particle-like wave packets that arise
in a wide range of physical contexts from a balance between dispersive
spreading and nonlinear focusing.  One of the key phenomena that
differentiates nonlinear coherent structures such as solitons from
their linear counterparts is what happens when such structures
interact.  Soliton solutions of equations with very special
mathematical structure (integrability) have been shown to interact
elastically \cite{zabusky_interaction_1965} and can be attractive or
repulsive \cite{gordon_interaction_1983}.  In more general systems,
soliton interactions can be significantly more complex, exhibiting
fusion, fission, annihilation, or spiraling
\cite{stegeman1999optical,krasnov2012radiative}.  A relative phase between the solitons
plays a dominant role in determining the resulting interaction
behaviors.  
An additional interaction feature, $90^\circ$ scattering, has been
predicted for two-dimensional (2D) magnetic solitons
\cite{Piette1998,komineas2001scattering} and solitons in field
theories \cite{Axenides2000,manton2004topological}.  The recent
experimental observation of a magnetic droplet soliton in a spatially
extended film \cite{Mohseni2013} provides the impetus for our deeper
study of magnetic soliton interactions.  Here, we show that the
interaction of a pair of 2D magnetic droplet solitons (from here on
in, droplets) exhibits rich behavior, principally dependent on the
droplets' relative phase.

Previous studies of soliton interaction in 2D ferromagnetic materials
have concentrated primarily on vortices, topological structures that
exhibit restricted dynamics \cite{papanicolaou1991dynamics}.  Unless
the ferromagnet is confined \cite{kasai_current-driven_2006},
conservation of overall topological charge pins the magnetic ``center
of mass'' in place, e.g. a single vortex core, limiting motion to
rotating collections \cite{Piette1998,komineas_rotating_2007} or
linear motion of vortex pairs with net zero topological charge.
Perpendicular scattering of two interacting vortex pairs has been
theoretically demonstrated \cite{komineas2001scattering}.  It appears
that $90^\circ$ scattering has a more universal character
\cite{manton2004topological}, not requiring a topological charge, and
previous numerical studies have indeed shown perpendicular scattering
even for approximate nontopological solitons \cite{Piette1998}.
Loosening topological restrictions and the fact that droplets, due to
their precessional nature, possess an extra degree of freedom (phase)
opens up many fascinating modes of interaction.

In this work, we classify head-on and angled droplet interactions in
terms of the droplets' relative phase and momenta via micromagnetic
simulations.  Sufficiently in-phase droplets experience an attractive
interaction that results in either merger into a new breathing bound
state for low speeds, or a scattering event transferring droplet
motion to the axis of symmetry.  Out-of-phase droplets experience a
repelling interaction that results in a scattering event obeying the
law of reflection.  Via symmetry, these results show that a
ferromagnetic boundary with pinned (free) spins repels (attracts) a
single droplet.  In particular, this provides an explanation for the
existence of ``edge droplets'' theoretically predicted for a spin
torque driven, confined ferromagnet with a free spin boundary
\cite{iacocca2013confined}.  At an intermediate relative phase, the
colliding droplets exhibit an ``explosion'' into spin waves and the
spontaneous formation of a single, breathing droplet.  This
annihilation behavior mimics particle colliders in which high energy
particles are smashed into byproducts.

\begin{figure*}
  \subfloat{ \hspace{-14.75mm} \raisebox{-1.15mm}{\includegraphics{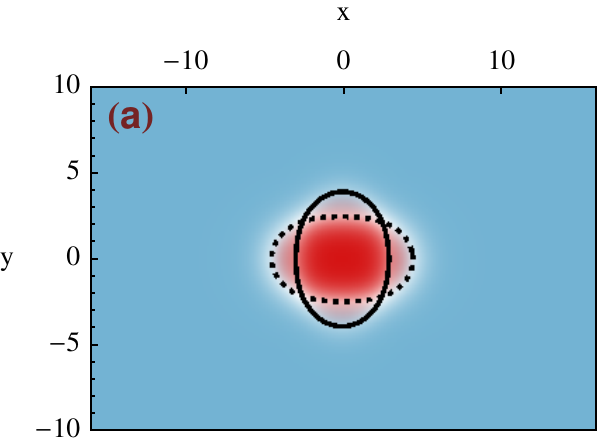}}}
  \subfloat{ \hspace{-1.2mm} \includegraphics{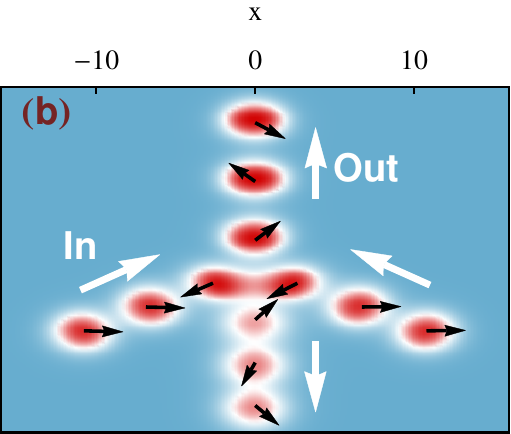}}
  \subfloat{\hspace{-0.2mm} \includegraphics{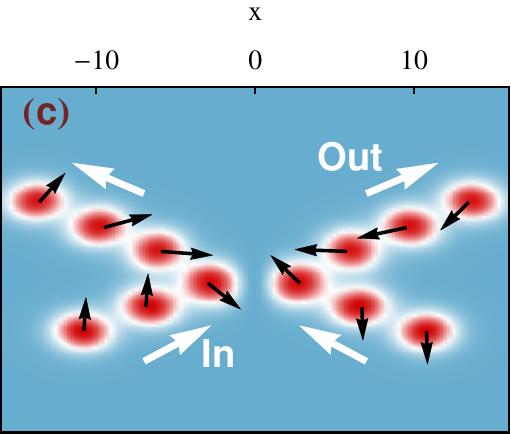}}\\[-1.1cm]
  \subfloat{\includegraphics{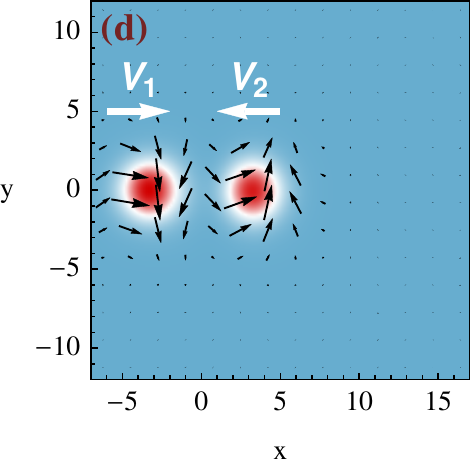}}
  \subfloat{ \hspace{-1mm} \includegraphics{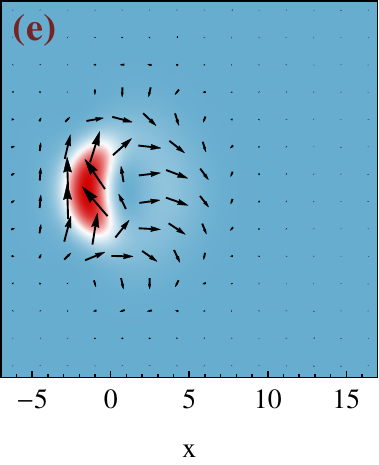}}
  \subfloat{\hspace{0mm} \includegraphics{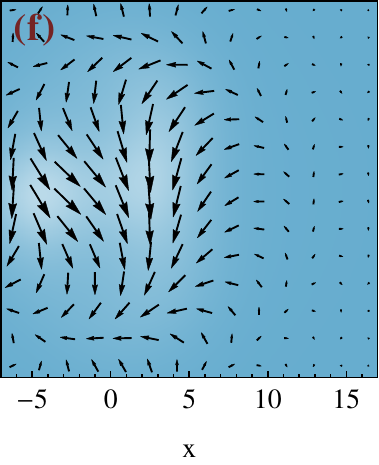}}
  \subfloat{\hspace{0mm} \includegraphics{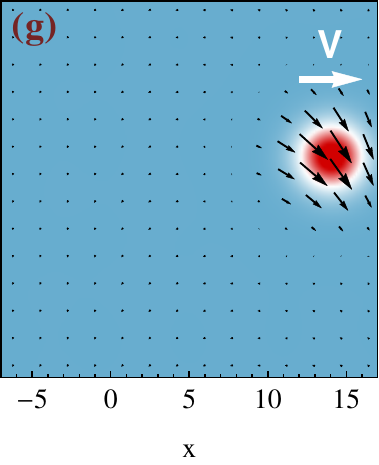}}
  \subfloat{\hspace{-1mm} \raisebox{3mm}{\includegraphics{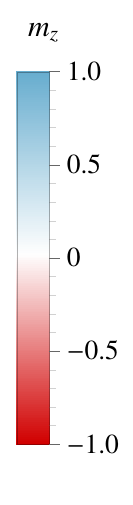}}}
  \caption{Droplet interactions.  (a) Breathing droplet at
    two times.  (b) In-phase merger and counterpropagation.  (c)
    Out-of-phase reflection.  (d-g) Droplet merger (e), annihilation
    to magnons (f), spontaneous breather formation (g).  (g inset)
    Spatial minimum of $m_z$ as a function of time for (d-g);
    vertical dashed lines denote times in (d-g).
  }
  \label{fig:interactions}
  \vspace{-5mm}
  \begin{picture}(0,0)
\put(112,68){{\color{black}\rule{2.63cm}{2.1cm}}}
\put(112.7,68.7){{\color{white}\rule{2.58cm}{2.05cm}}}
\put(113.78,72.2){\includegraphics{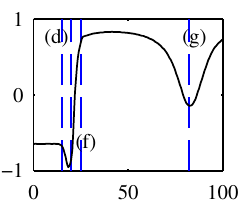}}
\put(113.7,69.7){{\color{white}\rule{24.5mm}{3mm}}}
\put(113.7,69.7){{\color{white}\rule{3mm}{20.1mm}}}
\put(143,69.7){\tiny Time}
\put(121.8,74.8){\tiny 0}
\put(147.25,74.8){\tiny 50}
\put(173.1,74.8){\tiny 100}
\put(117.1,79.3){\tiny -1}
\put(118.8,101.25){\tiny 0}
\put(118.8,123.1){\tiny 1}
\put(113.5,89){\rotatebox{90}{\tiny min($m_z$)}}
\end{picture}
\end{figure*}
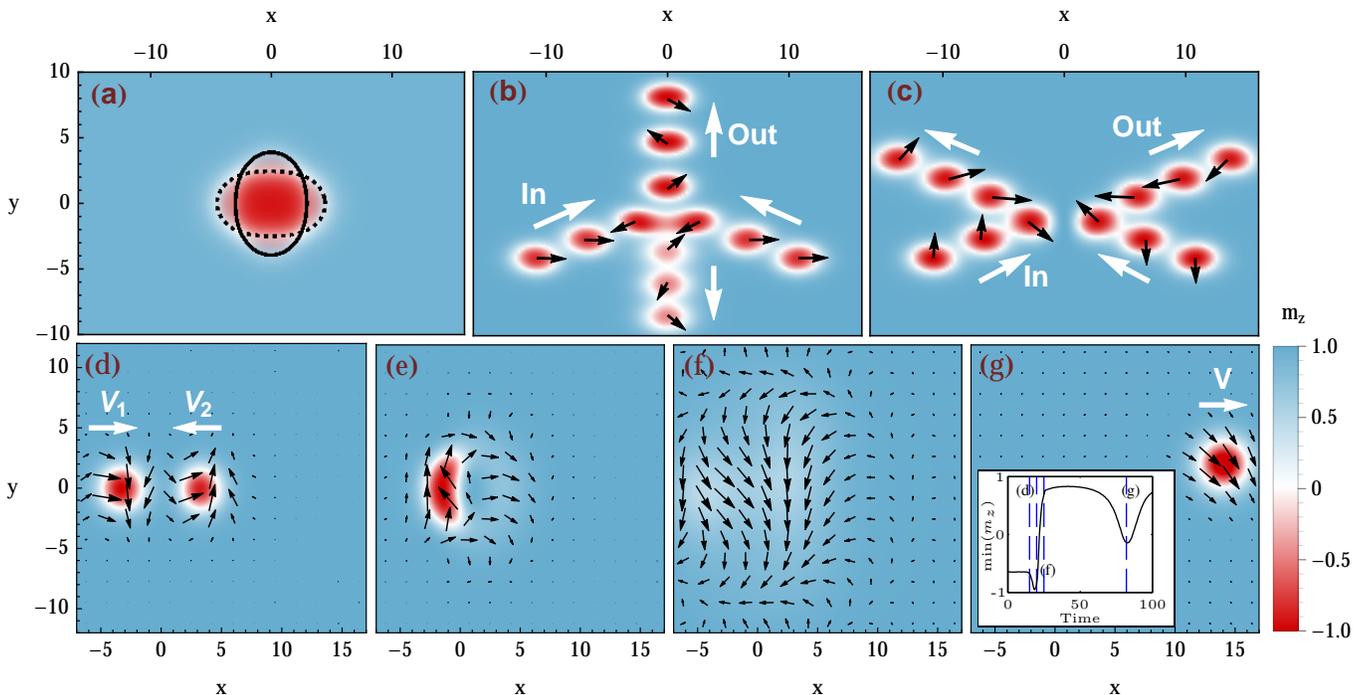
The model we consider is the Landau-Lifshitz torque equation with
perpendicular anisotropy,
\begin{equation}
  \label{eq:nondimLL}
  \begin{split}
    \frac{\partial \mathbf{m}}{\partial t} &= - \mathbf{m} \times
    [ \lap \mathbf{m} + (m_z + h_0)  \mathbf{z} ],
  \end{split}
\end{equation}
where $\lim_{\abs{\mathbf{x}}\rightarrow \infty} \mathbf{m} =
\mathbf{z}$.  Equation \eqref{eq:nondimLL} is an ultra-thin-film 2D
reduction of the full Landau-Lifshitz equation with long range
magnetostatic effects \cite{Hoefer2012}.  The magnetization vector is
normalized to unit length, spatial lengths are in units of $L_{\rm
  ex}/\sqrt{Q-1}$, times are in units of $[\abs{\gamma} \mu_0 M_s
(Q-1)]^{-1}$, and the perpendicular magnetic field magnitude $h_0 > 0$
is scaled by $M_{\rm s}(Q-1)$ where $L_{\rm ex}$ is the exchange
length, $\gamma$ is the gyromagnetic ratio, $\mu_0$ is the free space
permeability, $M_{\rm s}$ is the saturation magnetization, $Q = 2
K_u/(\mu_0 M_s^2) $ is the dimensionless quality factor, and $K_u$ is
the crystalline perpendicular anisotropy constant.  Here it is assumed
that $Q>1$ or equivalently that the perpendicular anisotropy is
sufficiently strong that it overcomes the effective planar anisotropy
due to the magnetostatic field.  This assumption is not an excessive
restriction as ferromagnets with this property such as CoFeB or Co/Ni
multilayers are currently in use (cf.~\cite{Mohseni2013}).  The energy
$\mathcal{E} = \frac{1}{2} \int [|\nabla \mathbf{m}|^2 + 1 - m_z^2 +
h_0(1-m_z)] d\mathbf{x}$ is conserved by solutions of
\eqref{eq:nondimLL}.  The magnetic field induces a positive shift of
precession frequency.  By entering the rotating frame, we take $h_0 =
0$ without loss of generality.

Droplet solutions of eq.~\eqref{eq:nondimLL} are parameterized by six
distinct quantities: initial position $\mathbf{x}^{0}$, initial
central phase $\phi^0$, propagation velocity $\mathbf{V}$, and rest
precession frequency $\omega$ \cite{Hoefer2012}.  A previous droplet
interaction study was limited to accurately computed stationary
(radially symmetric) droplets \cite{Piette1998}.  These solutions were
artificially deformed to induce propagation with a fixed, but not
prescribed speed and were accompanied by radiation.  Only in-phase,
head-on, approximate droplet interactions were considered.  In this
work, we leverage translation, rotation, and phase invariance of
eq.~\eqref{eq:nondimLL} in combination with a very accurate database
of precomputed propagating droplets \cite{Hoefer2012} in order to
explore a wide range of two-droplet initial conditions, each droplet
parameterized by $(\mathbf{x}^0_i, \phi^0_i,\mathbf{V}_i, \omega_i)$,
$i=1,2$.  The angle of interaction $\psi$ is the angle between
$\mathbf{V}_1$ and $\mathbf{V}_2$.  See \cite{numerics} for
micromagnetic details.

All the interactions described here depend principally on the relative
phase $\Delta = \phi_1 - \phi_2$ of the two initial droplets.  We find
that the interaction can be broadly classified as attractive or
repulsive with maximal attraction when $\Delta = 0$ varying to maximal
repulsion when $|\Delta| = \pi$, much as is the case for optical
solitons \cite{stegeman1999optical}, demonstrating the universality of
this behavior.  There is a critical, crossover phase $\Delta_{\rm cr}
> 0$ that divides the attractive and repulsive regimes.  Within this
general classification, there are four modes of interaction depending
on $\Delta$ and $\mathbf{V}_{1,2}$.  Figure \ref{fig:interactions}(a)
(small $V_i$, $|\Delta| < \Delta_{\rm cr}$): merger of two droplets
into a bound state whose perimeter is modulated (``breathes'') with
twice the precessional frequency.  Figure \ref{fig:interactions}(b)
($V_i$ large enough, $|\Delta| < \Delta_{\rm cr}$): merger followed by
counterpropagating droplets trapped along the axis of symmetry defined
by $\mathbf{V}_1 + \mathbf{V}_2$.  Figure \ref{fig:interactions}(c)
(any $V_i$, $\Delta_{\rm cr} < |\Delta| \le \pi$): reflection off the
symmetry axis.  Figures \ref{fig:interactions}(d-g) ($V_i$ large
enough, $|\Delta| \approx \Delta_{\rm cr}$): droplet merger and
annihilation into spin waves and a single propagating breather soliton.
Animations of all cases are available \cite{supplement}.  We now
describe each interaction category.

First, we consider the interaction of two stationary droplets,
$\mathbf{V}_i = 0$, initially situated so they weakly interact (10
units apart).  The initial droplets have the same frequency $\omega_1
= \omega_2 = \omega$, but varying relative phase.  For $\Delta_{\rm
  cr} < |\Delta| \le \pi$, the droplets slowly propagate away from
one another, exhibiting weak repulsion.  For $|\Delta| < \Delta_{\rm
  cr}$, the attraction interaction results in merger and then
perpendicular scattering.  Lacking sufficient momentum to overcome the
attraction, this merge-scatter process occurs many times, each with a
small loss of energy in the form of radiating spin waves until the
structure stabilizes into a breather state.  This two-droplet bound
state exhibits two frequencies: a precessional and a breathing
frequency, twice that of the precessional, at which the shape of the
new structure oscillates.  We have checked the numerically stable
evolution of the breather in Fig.~\ref{fig:interactions}(a) by
evolving it for 1400 time units.  For initial droplet frequencies
$\omega = 0.4$, the resulting new structure has precession frequency
$0.3$ and exhibits a deformation of shape as in the quarter-period
oscillation between the two configurations depicted in
Fig.~\ref{fig:interactions}(a). This new solitary wave is distinctly
different from the stationary droplet and what was observed in the
previous numerical study \cite{Piette1998} where, for $\Delta = 0$,
the two droplets were observed to merge-scatter, radiate spin waves,
and settle to a new, pure droplet with a single frequency.  This
merging behavior is similar to soliton fusion observed in optics
\cite{tikhonenko_three_1996}.

The next class of interactions we investigate are propagating droplets
with equal frequency $\omega_1 = \omega_2 = \omega$, equal speed
$V_1=V_2 = V$, and velocities reflected, $V_{1,x} = -V_{2,x}$, about
the $y$ axis so that $\mathbf{y}$ represents the axis of symmetry.
When the angle of interaction $\psi = \pi$, the collision is head-on.
The attractive interaction $|\Delta| < \Delta_{\rm cr}$ leads to
merger and ``trapped'' scattering along the $y$ axis as in
Fig.~\ref{fig:interactions}(c).  For the symmetric case when $V_{i,y}
= 0$, the scattering is $90^\circ$.  For the repelling interaction,
$\Delta_{\rm cr} < |\Delta| \le \pi$, the droplets reflect at an angle
equal to the angle of incidence $\psi/2$, as in
Fig.~\ref{fig:interactions}(b).  Both
Figs.~\ref{fig:interactions}(b,c) have $\omega = 0.4$, $\psi =
2\pi/3$, $V=0.6$, and successive plotted droplets are $t=10$ units
apart.  As $|\Delta|$ approaches $\Delta_{\rm cr}$, the two droplets
collide with one preferentially absorbing the other, transferring a
significant portion of their energy into spin waves followed by the
spontaneous formation of a breather state as shown in the head-on
collision of Figs.~\ref{fig:interactions}(d-g) with $\Delta =
\Delta_{\rm cr} = 92^\circ$, $\omega = 0.4$, $V = 0.6$, and $t \in
(30,40,80,164)$.  The asymmetry in the interaction of
Fig.~\ref{fig:interactions}(e-g) is due to the choice $0< \Delta <
\pi$.  A change in the sign of $\Delta$ reverses the asymmetry.
Figure \ref{fig:interactions}(g inset) demonstrates a steep depletion
of the excitation amplitude $1-m_z$ during the loss of energy to spin
waves and an amplitude coalescence associated with the formation of
the breather.  Annihilation therefore represents the crossover from
attractive to repulsive scattering where the incommensurate phases of
the colliding droplets cannot be resolved at high kinetic energies,
resulting in the explosive release of spin waves accompanied by
breather bound state formation.

Previous observations of soliton annihilation in optics were of a very
different type \cite{krolikowski_annihilation_1998} where the
simultaneous collision of three solitons could result in annihilation
of only one of them.  Here we see interaction behavior reminiscent of
high energy particles in a collider.  The byproducts of droplet
collision are a shower of magnons (spin waves) and a localized
breather.  Because a single droplet can be interpreted as a bound
state of magnon quasi-particles \cite{Kosevich1990}, the annihilation
interaction results in the irretrievable loss of energy to fundamental
constituents and a conglomerate state.

\begin{figure}
  \subfloat{ \hspace{-6mm}\includegraphics{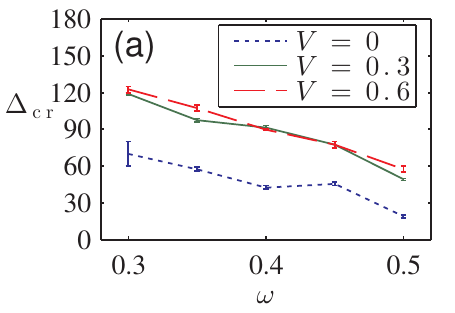}}
  \subfloat{\hspace{0mm} \includegraphics{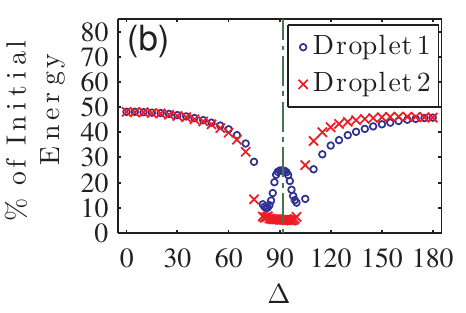}}\\[-6mm]
  \subfloat{ \hspace{-3mm}\includegraphics{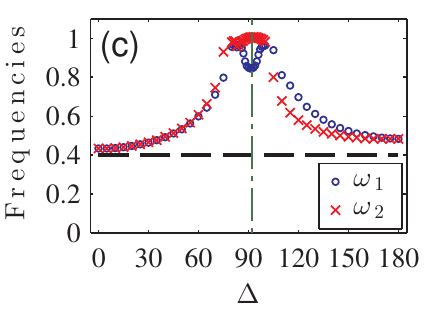}}
\subfloat{ \hspace{0mm} \includegraphics{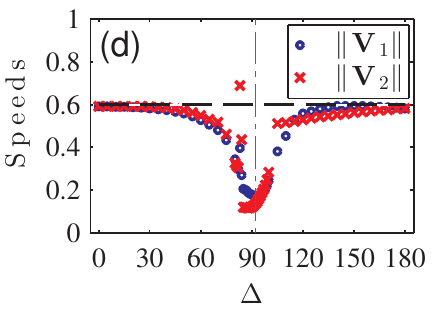}}
\caption{Head-on collision properties.  (a) Crossover phase for
  varying initial $\omega$, $V$.  (b-c) Post-collision properties for
  initial $\omega = 0.4$, $V = 0.6$.  Scattered
  droplet energy (b), frequency (c), and speed (d).}
  \label{fig:2}
  \vspace{-3mm}
\end{figure}


Now we investigate the interaction classification as both the initial
frequency $\omega = \omega_1 = \omega_2$ and velocities
$\mathbf{V}_{1} = (V,0) = -\mathbf{V}_2$ are varied for the head-on
collision configuration.  Figure \ref{fig:2}(a) depicts the
variation in the attractive to repulsive crossover parameter
$\Delta_{\rm cr}$.  Generally, for any initial speed $V$, $\Delta_{\rm
  cr}$ decreases with increasing $\omega$ showing that the repulsive
interaction is favored for smaller amplitude droplets (increasing
$\omega$, $V$ lead to a decrease in droplet amplitude
\cite{Hoefer2012}).  Colliding droplets with $V \in \{0.3,0.6\}$
exhibit approximately the same crossover, in contrast to $V = 0$,
which is downshifted by about $30^\circ$-$40^\circ$.  We deduce that
propagating droplets favor attractive scattering across a wider range
of phases than initially stationary droplets.  Moving droplets exhibit
an underlying spin-wave-type structure with wavenumber $k = V/2$ that
is associated with nonzero local topological density
\cite{Piette1998,Hoefer2012}, whereas stationary droplets have a
uniform phase and zero topological density \cite{Kosevich1990}.  We
never observed stationary droplets to annihilate so the difference in
$\Delta_{\rm cr}$ for moving and stationary droplets and the existence
of annihilation may be attributable to the complexity introduced by
nonzero $k$ and topological density associated with propagating
droplets.

When $0 < |\Delta| < \Delta_{\rm cr}$, the result of attractive
interaction is droplets of different speeds and frequencies relative
to their initial values.  The post-interaction droplet properties for
head-on collisions with varying $\Delta$ are shown in
Figs.~\ref{fig:2}(b-d).  Generically, in-phase and out-of-phase
interactions exhibit symmetric behavior with the least energy loss to
radiation. For most $\Delta$ below $\Delta_{\rm cr}$, the post droplet
frequencies and speeds are roughly symmetric.  Near and above the
crossover $\Delta_{\rm cr}$, however, there is significant asymmetric
energy loss due to increased spin wave radiation and energy exchange
between the droplets with their frequencies and speeds approaching the
linear spin wave band $\omega = 1 - V^2/4$ \cite{Hoefer2012}.  As
$\Delta$ is increased, Fig.~\ref{fig:2}(b) shows the energy retained
in the post-scattered droplets decreasing until it reaches a minimum.
Above this value of $\Delta$ we begin to observe the annihilation
interaction.  The local extremum (at $92^\circ$ for the parameters in
Fig.~\ref{fig:2}(b)) serves as the definition of $\Delta_{\rm cr}$.
When $\Delta_{\rm cr} < \Delta < \pi$, the post-interaction droplet
with greater energy is reflected to the right, the roles reversed if
the sign of $\Delta$ is changed.

For in-phase droplets not propagating head-on, we observe droplet
scattering along the direction $\mathbf{V}_1+\mathbf{V}_2$.  The
asymmetry in energy transfer post interaction is accentuated in
Fig.~\ref{fig:interactions}(b).  This asymmetry is due to the
conserved positive momentum in the direction
$\mathbf{V}_1+\mathbf{V}_2$, favoring larger droplets. For small
$\psi$, the collision results in approximately a single droplet. This
behavior varies in a continuous fashion, limiting to the case when
$\psi = \pi$ for $90^\circ$ scattering of droplets with equal size and
frequency.

The model \eqref{eq:nondimLL} neglects several important physical
effects. For example, relaxation processes (damping) in ferromagnets
are typically weak but play an important role in experiments and
soliton dynamics \cite{baryakhtar1997}.  Long range magnetostatics
affects any ferromagnet with finite thickness.  We numerically
investigate the impact on droplet interactions due to Landau-Lifshitz
damping with damping parameter $\alpha = 0.01$ and a 2D, thickness
dependent correction to the magnetostatic field
\cite{garcia-cervera_one-dimensional_2004,bookman2013analytical} with
thickness parameter $\delta = 0.5$.  We did not observe a significant
qualitative change in the resulting numerical experiments. As observed
previously, the effect of damping is to cause droplets to accelerate
while the frequency increases
\cite{baryakhtar1997,hoefer2012propagation}.  Magnetostatics result in
a negative frequency shift of the droplet
\cite{bookman2013analytical}.  We find that the quantitative locations
of the breathing, counterpropagation, reflection, and annihilation
regions are changed under these perturbations.  Nevertheless, we still
observe all four phenomena over sufficiently short time scales so that
damping has not completely relaxed the magnet to equilibrium.

Finally, we investigate droplet collisions with different frequencies
and velocities.  Droplet pairings are chosen so that their frequencies
and velocities differed but their momenta did not.  This has a
significant impact on the post-collision frequencies and
velocities. The four interaction categories are all observed. Common
interaction behaviors are reflection or annihilation resulting in a
breather propagating in the direction of overall momentum.

\begin{figure}
  \centering
  \subfloat{\includegraphics{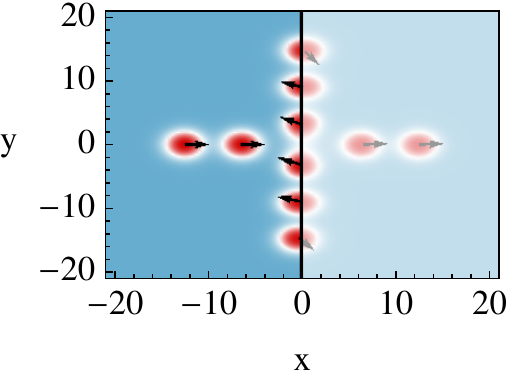}
    \subfloat{\hspace{0mm} \includegraphics{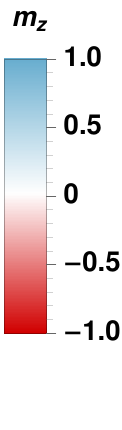}}
    \hspace{1mm}} 
  \caption{Method of images depicting head-on collision of
    droplet with a free spin boundary
    (vertical line).  Two counterpropagating edge droplets are
    created.}
  \label{fig:edge}
\end{figure}
While the case of two initial droplets with the same speed and
frequency may seem restrictive, it is highly relevant in applications.
Real ferromagnets can exhibit boundaries with either pinned
($\mathbf{m} = \mathbf{z}$) or free ($\partial \mathbf{m}/\partial
\mathbf{n} = 0$, $\mathbf{n}$ a boundary normal) spins.  We can
utilize symmetries of the droplet solution \cite{Hoefer2012} and of
eq.~\eqref{eq:nondimLL} in order to implement a method of images
whereby we reflect an initial droplet about the $y$ axis, taking $V_x
\to - V_x$.  The choice of two in-phase droplets corresponds to an
even reflection and a free spin boundary condition along the $y$ axis.
The choice of two out-of-phase droplets corresponds to an odd
reflection of $(m_x,m_y)$ and an even reflection of $m_z$ leading to a
pinned spin boundary condition.  Thus, the entire discussion of
interacting droplets with $\Delta = 0$ or $\Delta = \pi$ (e.g.,
Figs.~\ref{fig:interactions}(a-c)) translates to droplet scattering
off a boundary.  This is directly illustrated in Fig.~\ref{fig:edge}
where an in-phase, head-on collision with $(\omega,V) = (0.4,0.6)$
results in $90^\circ$ scattering and the generation of two edge
droplets counterpropagating along the $y$ axis.  The free spin
boundary therefore attracts droplets.  This has been observed in
micromagnetic simulations of ferromagnetic nanowires with a spin
torque nanocontact (NC) \cite{iacocca2013confined}.  The attractive
force of the free spin boundary overcomes the restoring force of the
NC \cite{bookman2013analytical} resulting in an edge droplet for
sufficiently narrow wires.  Since out-of-phase droplets repel one
another, the pinned spin boundary repels droplets. This suggests a way
to create a droplet waveguide in a nanowire.  If both edges of the
nanowire are pinned with vertical magnetization, the droplet is
repelled from the boundary.  This observation coupled with the ability
to accelerate droplets with magnetic field gradients
\cite{hoefer2012propagation} suggests a practical method to stably
propagate droplets in patterned media.

In summary, we have classified the interactions of two magnetic
droplet solitons into four types depending on their relative phase and
speed, observing a new nontopological structure, the droplet
breather, and demonstrating attractive, repulsive, and annihilation
behaviors.

\begin{acknowledgments}
  The authors gratefully acknowledge support from NSF CAREER grant
  DMS-1255422.
\end{acknowledgments}


\bibliographystyle{apsrev4-1}

%

\end{document}